\documentclass[a4paper,11pt]{article}
\usepackage{pos}

\usepackage[symbol]{footmisc}
\usepackage[english]{babel}
\usepackage{graphicx}
\usepackage{graphics}
\usepackage{braket}
\usepackage{bbold}
\usepackage{amsmath}
\usepackage{nicefrac}
\usepackage{dcolumn}% Align table columns on decimal point
\usepackage{bm}% bold math
\usepackage{slashed}
\usepackage{datetime}
\usepackage{mciteplus}
\usepackage{multirow}
\usepackage{bbold}
\usepackage{siunitx}
\usepackage{booktabs}
\usepackage{color, soul}
\usepackage[usenames,dvipsnames]{xcolor}
\usepackage{float}
\usepackage[utf8]{inputenc}
\usepackage[normalem]{ulem}
\usepackage{mathtools}
\usepackage{setspace}

\renewcommand{\thefootnote}{\fnsymbol{footnote}}

\newcommand{\LQCD}{\Lambda_{\rm QCD}}

\newcommand{\NLLm}{{\rm NLL/NLO^-}}

\newcommand{\DY}{\Delta Y}

\title{Towards Higgs and $Z$ boson plus jet distributions \\ at NLL/NLO$^+$}
\ShortTitle{Towards Higgs and $Z$ boson plus jet distributions at NLL/NLO$^+$}

\author*[a]{Francesco Giovanni Celiberto}
\author[b,c]{Luigi Delle Rose}
\author[d]{Michael Fucilla}
\author[b,c]{\\Gabriele Gatto}
\author[b,c]{Alessandro Papa}

\affiliation[a]{Universidad de Alcalá (UAH), Departamento de Física y Matemáticas, Campus Universitario, \\ Alcalá de Henares, E-28805, Madrid, Spain}

\affiliation[b]{Dipartimento di Fisica, Università della Calabria, Arcavacata di Rende, I-87036, Cosenza, Italy}

\affiliation[c]{
INFN, Gruppo Collegato di Cosenza, Arcavacata di Rende, I-87036, Cosenza, Italy}

\affiliation[d]{Université Paris-Saclay, CNRS/IN2P3, IJCLab, 91405, Orsay, France}

\emailAdd{francesco.celiberto@uah.es}
\emailAdd{luigi.dellerose@unical.it}
\emailAdd{michael.fucilla@ijclab.in2p3.fr}
\emailAdd{gabriele.gatto@unical.it}
\emailAdd{alessandro.papa@fis.unical.it}

\abstract{We present novel predictions for rapidity and transverse-momentum distributions sensitive to the emission of a Higgs boson accompanied by a jet in proton collisions, calculated within the NLO fixed order in QCD and matched with the next-to leading energy-logarithmic accuracy. 
We also highlight first advancements in the extension of our analysis to the $Z$-boson case. We come out with the message that the improvement of fixed-order calculations on Higgs- and $Z$-boson plus jet distributions is a required step to reach the precision level of the description of observables relevant for Higgs and electroweak physics at current LHC energies and nominal FCC ones.}

\FullConference{31st International Workshop on Deep Inelastic Scattering (DIS2024)\\
 8–12 April 2024\\
Grenoble, France}

\begin{document}
\maketitle

%%%----------------------------------------
\section{Introduction}
\label{sec:introduction}
%%%----------------------------------------

To properly describe Higgs- and $Z$-boson production rates at the LHC as well as the future FCC, the \emph{all-order} resummation of energy logarithms is relevant.
In this study we turn our attention to the \emph{semi-hard} sector, where the stringent scale hierarchy, $\LQCD \ll \{Q_i\} \ll \sqrt{s}$, with $\{Q_i\}$ a set of typical hard scales and $\sqrt{s}$ the center-of-mass energy, leads to large energy logarithms.
The Balitsky--Fadin--Kuraev--Lipatov (BFKL) resummation~\cite{Fadin:1975cb,Balitsky:1978ic} accounts for these logarithms within the leading-logarithmic (LL) and next-to-leading logarithmic (NLL) order.
It also permits us to access the low-$x$ gluon density in the proton~\cite{Bacchetta:2020vty,Bacchetta:2024fci,Bacchetta:2021lvw,Bacchetta:2021twk,Bacchetta:2024uxb,Arbuzov:2020cqg,Celiberto:2021zww,Amoroso:2022eow,Bolognino:2018rhb,Bolognino:2021niq,Hentschinski:2022xnd,Celiberto:2019slj}.
Excellent probes of high-energy QCD in proton collisions are semi-inclusive productions of two particles tagged with high transverse masses and a strong rapidity separation, $\DY$.
To describe these two-particle reactions, a \emph{multilateral} approach, where both collinear and high-energy dynamics are embodied, needs to be used.
To this end, a \emph{hybrid} factorization formalism (HyF) was built~\cite{Celiberto:2020tmb,Bolognino:2021mrc} (see also~\cite{vanHameren:2022mtk,Bonvini:2018ixe,Silvetti:2022hyc} for single-particle emissions).
HyF cross sections read as convolutions of two reaction-dependent emission functions and a universal NLL BFKL Green's function, which corresponds to the Sudakov radiator of soft-gluon resummations.
Emission functions are in turn factorized as a convolution of collinear parton densities (PDFs) and singly off-shell coefficient functions.
The highest accuracy of HyF is NLL/NLO: for a given reaction, the corresponding coefficient functions need to be calculated at fixed next-to-leading order (NLO) accuracy. 
Contrarily, one must rely on a partial next-to-leading level ($\NLLm$), with the Green's function taken at NLL, one coefficient function at NLO, and the other one at LO.
The HyF formalism has been tested so far \emph{via}: Mueller--Navelet jet tags~\cite{Ducloue:2013hia,Ducloue:2013bva,Celiberto:2015yba,Celiberto:2015mpa,Celiberto:2016ygs,Celiberto:2017ius,Caporale:2018qnm,Celiberto:2022gji,Egorov:2023duz,Baldenegro:2024ndr}, Drell--Yan pair~\cite{Celiberto:2018muu,Golec-Biernat:2018kem}, light~\cite{Celiberto:2016hae,Celiberto:2017ptm,Bolognino:2018oth,Bolognino:2019yqj,Bolognino:2019cac,Celiberto:2020rxb,Celiberto:2022kxx} as well as heavy-light~\cite{Celiberto:2017nyx,Bolognino:2019yls,Bolognino:2019ccd,AlexanderAryshev:2022pkx,Celiberto:2021dzy,Celiberto:2021fdp,Celiberto:2022zdg,Celiberto:2022keu,Celiberto:2024omj,Anchordoqui:2021ghd,Feng:2022inv} hadron, quarkonium~\cite{Boussarie:2017oae,Chapon:2020heu,Celiberto:2022dyf,Celiberto:2023fzz}, and exotic-matter~\cite{Celiberto:2023rzw,Celiberto:2024mab,Celiberto:2024mrq} detections.
Here we study Higgs-plus-jet rates~\cite{Celiberto:2020tmb,Celiberto:2023rtu}, which have been already investigated at next-to-NLO perturbative QCD~\cite{Chen:2014gva,Boughezal:2015dra,Dawson:2022zbb} and by the next-to-NLL transverse-momentum resummation~\cite{Monni:2019yyr}.
We present the POWHEG+JETHAD method, a novel procedure aimed at \emph{matching} the NLO fixed-order with the NLL high-energy resummation.

%%%----------------------------------------
\section{Towards Higgs-plus-jet production at NLL/NLO}
\label{sec:matching}
%%%----------------------------------------

\begin{figure*}[!t]
\centering

\includegraphics[scale=0.36,clip]{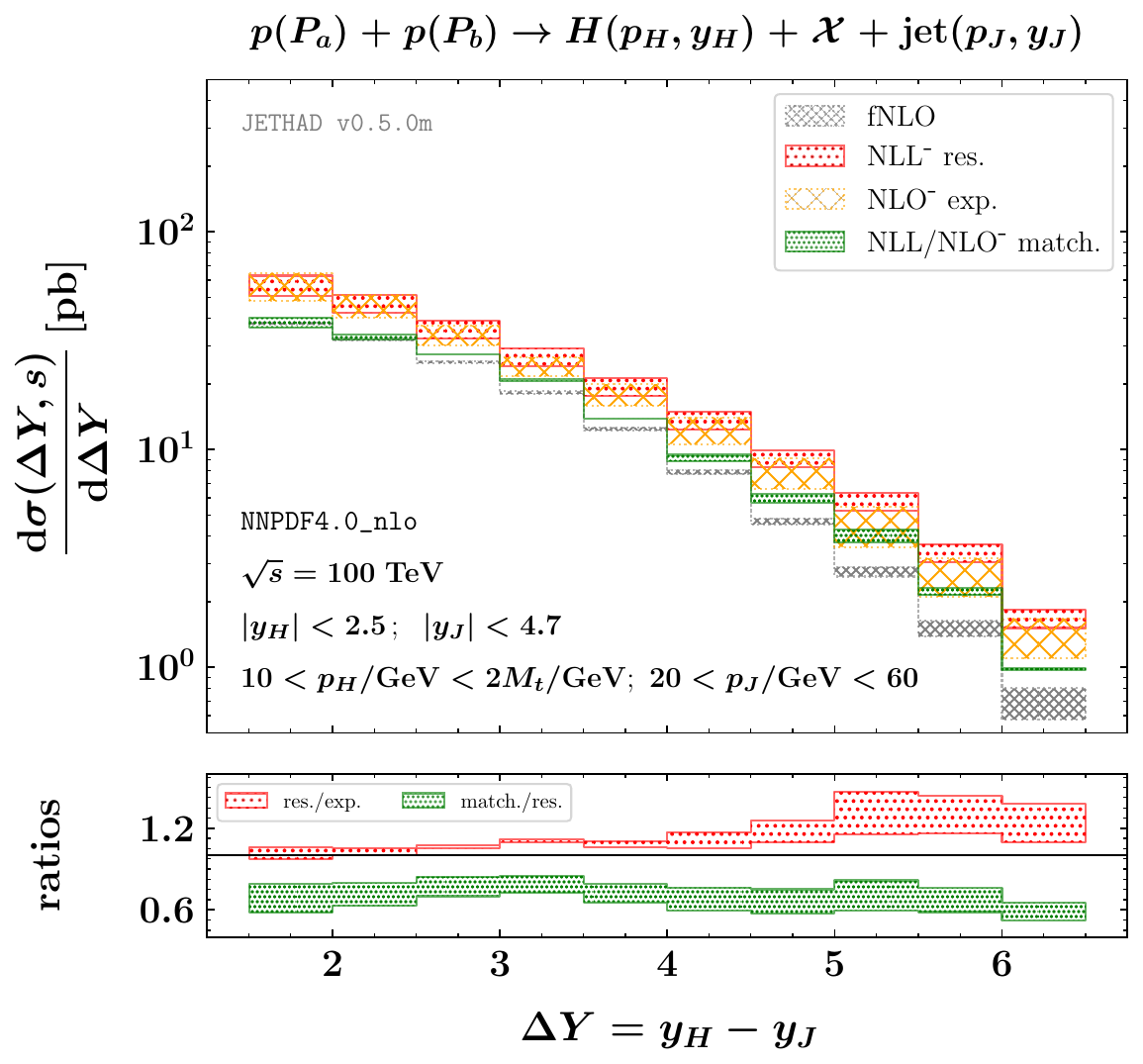}
 \hspace{0.30cm}
\includegraphics[scale=0.36,clip]{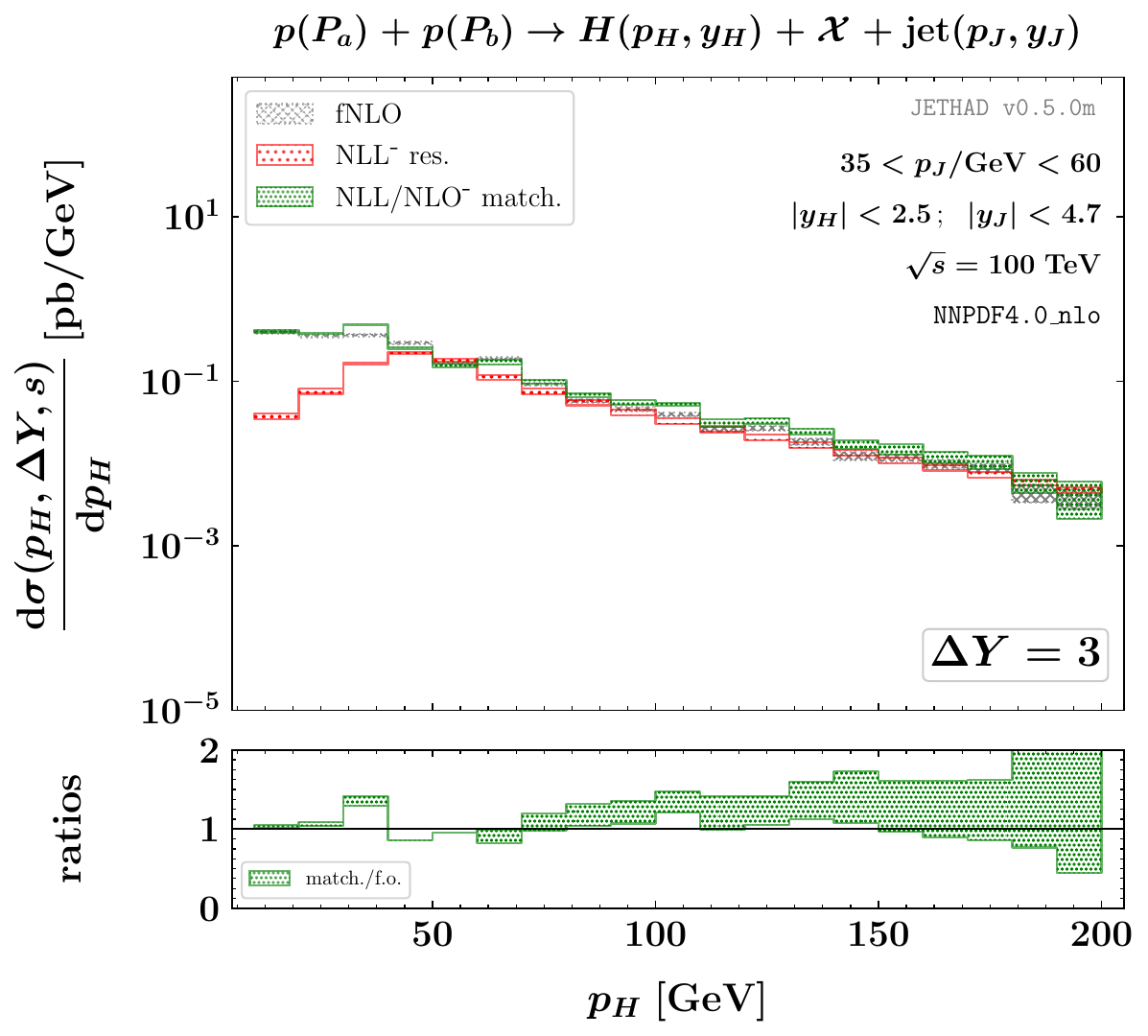}

\caption{Higgs-plus-jet $\DY$ (left) and $p_H$ spectrum at $100$ TeV nominal FCC energies.
Uncertainty bands show $\mu_{R,F}$ variation in the $1 < C_{\mu} < 2$ range. Text boxes refer to kinematic cuts.
}

\label{fig:predictions}
\end{figure*}

Preliminary HyF studies on the Higgs transverse-momentum ($p_H$) spectrum in the inclusive Higgs-plus-jet channel at LHC~\cite{Celiberto:2020tmb} and FCC~\cite{Celiberto:2023rtu} energies exhibited a solid stability under radiative corrections and scale variations.
Nevertheless, a strong discrepancy between HyF predictions and the pure fixed-order background arose.
For this reason, we developed a prime \emph{matching} method between the NLO fixed order and the high-energy NLL resummation.
It bases upon the exact removal, at the $\NLLm$ accuracy, of the corresponding \emph{double counting}.
Because the NLO Higgs emission function~\cite{Hentschinski:2020tbi,Celiberto:2022fgx,Nefedov:2019mrg} still has to be implemented in the JETHAD code~\cite{Celiberto:2020wpk,Celiberto:2022rfj,Celiberto:2023fzz,Celiberto:2024mrq,Celiberto:2024swu}, we will rely upon a $\NLLm$ description.
Our matching procedure reads as follows~\cite{Celiberto:2023uuk_article,Celiberto:2023eba_article,Celiberto:2023nym}

\begin{equation}
\label{eq:matching}
\begin{split}
 \hspace{-0.155cm}
 \underbrace{{\rm d}\sigma^{{{\rm NLL/NLO}}^{\boldsymbol{-}}}(\Delta Y, \varphi, s)}_{\text{\colorbox{OliveGreen}{\textbf{\textcolor{white}{NLL/NLO$^{\boldsymbol{-}}$}}} {\tt POWHEG+JETHAD}}} 
 = 
 \underbrace{{\rm d}\sigma^{\rm NLO}(\Delta Y, \varphi, s)}_{\text{\colorbox{gray}{\textcolor{white}{\textbf{NLO}}} {\tt POWHEG} w/o PS}}
 +\; 
 \underbrace{\underbrace{{\rm d}\sigma^{{{\rm NLL}}^{\boldsymbol{-}}}(\Delta Y, \varphi, s)}_{\text{\colorbox{red}{\textbf{\textcolor{white}{NLL$^{\boldsymbol{-}}$ resum}}} (HyF)}}
 \;-\; 
 \underbrace{\Delta{\rm d}\sigma^{{{\rm NLL/NLO}}^{\boldsymbol{-}}}(\Delta Y, \varphi, s)}_{\text{\colorbox{orange}{\textbf{NLL$^{\boldsymbol{-}}$ expanded}} at NLO}}}_{\text{\colorbox{NavyBlue}{\textbf{\textcolor{white}{NLL$^{\boldsymbol{-}}$}}} {\tt JETHAD} w/o NLO$^{\boldsymbol{-}}$ double counting}}
 \,.
\end{split}
\end{equation}

An observable, ${\rm d}\sigma^{{{\rm NLL/NLO}}^{\boldsymbol{-}}}$, matched at $\NLLm$ (green) by the POWHEG+JETHAD method, is cast as a sum of the NLO fixed order (gray) from POWHEG~\cite{Hamilton:2012rf,Bagnaschi:2023rbx,Banfi:2023mhz} (without \emph{parton shower} (PS)~\cite{Alioli:2022dkj,vanBeekveld:2022zhl,FerrarioRavasio:2023kyg}) and the $\rm NLL^-$ resummed part (blue) from JETHAD. The latter represents the $\rm NLL^-$ HyF resummed term (red) minus the $\rm NLL^-$ expanded one (orange) at NLO, \emph{i.e.} without double counting.
To extend our preliminary analysis presented in Refs.~\cite{Celiberto:2023uuk_article,Celiberto:2023eba_article,Celiberto:2023nym} we show 100~TeV~FCC predictions for the $\DY$ spectrum (Fig.~\ref{fig:predictions}, left panel) and the $p_H$ one at $\DY = 3$ (Fig.~\ref{fig:predictions}, right panel).

%%%----------------------------------------
\section{Conclusions and outlook}
\label{sec:conclusions}
%%%----------------------------------------

We proposed a new procedure, based on the POWHEG~\cite{Hamilton:2012rf,Bagnaschi:2023rbx,Banfi:2023mhz} and JETHAD~\cite{Celiberto:2020wpk,Celiberto:2022rfj,Celiberto:2023fzz,Celiberto:2024mrq,Celiberto:2024swu} codes, aimed at matching NLO fixed-order predictions with the NLL energy resummation and beyond (NLL/NLO$^+$).
Next steps will include: $a)$ NLO contributions to the Higgs emission function~\cite{Hentschinski:2020tbi,Celiberto:2022fgx,Nefedov:2019mrg}, $b)$ heavy-quark mass contributions~\cite{Jones:2018hbb,Bonciani:2022jmb}, $c)$ the extension to the $Z$-boson case.

% %%%----------------------------------------
% \section*{Acknowledgments}
% \label{sec:acknowledgments}
% %%%----------------------------------------

% This work was supported by the Atracci\'on de Talento Grant n. 2022-T1/TIC-24176 of the Comunidad Aut\'onoma de Madrid, Spain, and by the INFN/QFT@COLLIDERS Project, Italy. M.~F. is supported by Agence Nationale de la Recherche under the contract ANR-17-CE31-0019.

%-----------------------------------------
\begingroup
\setstretch{0.6}
\bibliographystyle{bibstyle}
\bibliography{biblography}

\begin{thebibliography}{78}
\expandafter\ifx\csname natexlab\endcsname\relax\def\natexlab#1{#1}\fi
\expandafter\ifx\csname bibnamefont\endcsname\relax
  \def\bibnamefont#1{#1}\fi
\expandafter\ifx\csname bibfnamefont\endcsname\relax
  \def\bibfnamefont#1{#1}\fi
\expandafter\ifx\csname citenamefont\endcsname\relax
  \def\citenamefont#1{#1}\fi
\expandafter\ifx\csname url\endcsname\relax
  \def\url#1{\texttt{#1}}\fi
\expandafter\ifx\csname urlprefix\endcsname\relax\def\urlprefix{URL }\fi
\providecommand{\bibinfo}[2]{#2}
\providecommand{\eprint}[2][]{\url{#2}}

\bibitem[{\citenamefont{Fadin et~al.}(1975)}]{Fadin:1975cb}
\bibinfo{author}{\bibfnamefont{V.~S.} \bibnamefont{Fadin}} \bibnamefont{et~al.}, \bibinfo{journal}{Phys. Lett. B} \textbf{\bibinfo{volume}{60}}, \bibinfo{pages}{50} (\bibinfo{year}{1975}).

\bibitem[{\citenamefont{Balitsky\mbox{, L. N. Lipatov}}(1978)}]{Balitsky:1978ic}
\bibinfo{author}{\bibfnamefont{I.~I.} \bibnamefont{Balitsky\mbox{, L. N. Lipatov}}}, \bibinfo{journal}{Sov.\ J.\ Nucl.\ Phys.} \textbf{\bibinfo{volume}{28}}, \bibinfo{pages}{822} (\bibinfo{year}{1978}).

\bibitem[{\citenamefont{Bacchetta et~al.}(2020)}]{Bacchetta:2020vty}
\bibinfo{author}{\bibfnamefont{A.}~\bibnamefont{Bacchetta}} \bibnamefont{et~al.}, \bibinfo{journal}{Eur. Phys. J. C} \textbf{\bibinfo{volume}{80}}, \bibinfo{pages}{733} (\bibinfo{year}{2020}), \eprint{2005.02288}.

\bibitem[{\citenamefont{Bacchetta et~al.}(2024{\natexlab{a}})\citenamefont{Bacchetta, Celiberto,  Radici}}]{Bacchetta:2024fci}
\bibinfo{author}{\bibfnamefont{A.}~\bibnamefont{Bacchetta}}, \bibinfo{author}{\bibfnamefont{F.~G.} \bibnamefont{Celiberto}},  \bibinfo{author}{\bibfnamefont{M.}~\bibnamefont{Radici}}, \bibinfo{journal}{Eur. Phys. J. C} \textbf{\bibinfo{volume}{84}}, \bibinfo{pages}{576} (\bibinfo{year}{2024}{\natexlab{a}}), \eprint{2402.17556}.

\bibitem[{\citenamefont{Bacchetta et~al.}(2022{\natexlab{a}})\citenamefont{Bacchetta, Celiberto,  Radici}}]{Bacchetta:2021lvw}
\bibinfo{author}{\bibfnamefont{A.}~\bibnamefont{Bacchetta}}, \bibinfo{author}{\bibfnamefont{F.~G.} \bibnamefont{Celiberto}},  \bibinfo{author}{\bibfnamefont{M.}~\bibnamefont{Radici}}, \bibinfo{journal}{PoS} \textbf{\bibinfo{volume}{EPS-HEP2021}}, \bibinfo{pages}{376} (\bibinfo{year}{2022}{\natexlab{a}}), \eprint{2111.01686}.

\bibitem[{\citenamefont{Bacchetta et~al.}(2022{\natexlab{b}})\citenamefont{Bacchetta, Celiberto,  Radici}}]{Bacchetta:2021twk}
\bibinfo{author}{\bibfnamefont{A.}~\bibnamefont{Bacchetta}}, \bibinfo{author}{\bibfnamefont{F.~G.} \bibnamefont{Celiberto}},  \bibinfo{author}{\bibfnamefont{M.}~\bibnamefont{Radici}}, \bibinfo{journal}{PoS} \textbf{\bibinfo{volume}{PANIC2021}}, \bibinfo{pages}{378} (\bibinfo{year}{2022}{\natexlab{b}}), \eprint{2111.03567}.

\bibitem[{\citenamefont{Bacchetta et~al.}(2024{\natexlab{b}})\citenamefont{Bacchetta, Celiberto,  Radici}}]{Bacchetta:2024uxb}
\bibinfo{author}{\bibfnamefont{A.}~\bibnamefont{Bacchetta}}, \bibinfo{author}{\bibfnamefont{F.~G.} \bibnamefont{Celiberto}},  \bibinfo{author}{\bibfnamefont{M.}~\bibnamefont{Radici}}, \bibinfo{journal}{PoS} \textbf{\bibinfo{volume}{SPIN2023}}, \bibinfo{pages}{049} (\bibinfo{year}{2024}{\natexlab{b}}), \eprint{2406.04893}.

\bibitem[{\citenamefont{Arbuzov et~al.}(2021)}]{Arbuzov:2020cqg}
\bibinfo{author}{\bibfnamefont{A.}~\bibnamefont{Arbuzov}} \bibnamefont{et~al.}, \bibinfo{journal}{Prog. Part. Nucl. Phys.} \textbf{\bibinfo{volume}{119}}, \bibinfo{pages}{103858} (\bibinfo{year}{2021}), \eprint{2011.15005}.

\bibitem[{\citenamefont{Celiberto}(2021{\natexlab{a}})}]{Celiberto:2021zww}
\bibinfo{author}{\bibfnamefont{F.~G.} \bibnamefont{Celiberto}}, \bibinfo{journal}{Nuovo Cim.} \textbf{\bibinfo{volume}{C44}}, \bibinfo{pages}{36} (\bibinfo{year}{2021}{\natexlab{a}}), \eprint{2101.04630}.

\bibitem[{\citenamefont{Amoroso et~al.}(2022)}]{Amoroso:2022eow}
\bibinfo{author}{\bibfnamefont{S.}~\bibnamefont{Amoroso}} \bibnamefont{et~al.}, \bibinfo{journal}{Acta Phys. Polon. B} \textbf{\bibinfo{volume}{53}}, \bibinfo{pages}{A1} (\bibinfo{year}{2022}), \eprint{2203.13923}.

\bibitem[{\citenamefont{Bolognino et~al.}(2018{\natexlab{a}})}]{Bolognino:2018rhb}
\bibinfo{author}{\bibfnamefont{A.~D.} \bibnamefont{Bolognino}} \bibnamefont{et~al.}, \bibinfo{journal}{Eur. Phys. J.} \textbf{\bibinfo{volume}{C78}}, \bibinfo{pages}{1023} (\bibinfo{year}{2018}{\natexlab{a}}), \eprint{1808.02395}.

\bibitem[{\citenamefont{Bolognino et~al.}(2021{\natexlab{a}})}]{Bolognino:2021niq}
\bibinfo{author}{\bibfnamefont{A.~D.} \bibnamefont{Bolognino}} \bibnamefont{et~al.}, \bibinfo{journal}{Eur. Phys. J. C} \textbf{\bibinfo{volume}{81}}, \bibinfo{pages}{846} (\bibinfo{year}{2021}{\natexlab{a}}), \eprint{2107.13415}.

\bibitem[{\citenamefont{Hentschinski et~al.}(2023)}]{Hentschinski:2022xnd}
\bibinfo{author}{\bibfnamefont{M.}~\bibnamefont{Hentschinski}} \bibnamefont{et~al.}, \bibinfo{journal}{Acta Phys. Polon. B} \textbf{\bibinfo{volume}{54}}, \bibinfo{pages}{2} (\bibinfo{year}{2023}), \eprint{2203.08129}.

\bibitem[{\citenamefont{Celiberto}(2019)}]{Celiberto:2019slj}
\bibinfo{author}{\bibfnamefont{F.~G.} \bibnamefont{Celiberto}}, \bibinfo{journal}{Nuovo Cim.} \textbf{\bibinfo{volume}{C42}}, \bibinfo{pages}{220} (\bibinfo{year}{2019}), \eprint{1912.11313}.

\bibitem[{\citenamefont{Celiberto et~al.}(2021{\natexlab{a}})}]{Celiberto:2020tmb}
\bibinfo{author}{\bibfnamefont{F.~G.} \bibnamefont{Celiberto}} \bibnamefont{et~al.}, \bibinfo{journal}{Eur. Phys. J. C} \textbf{\bibinfo{volume}{81}}, \bibinfo{pages}{293} (\bibinfo{year}{2021}{\natexlab{a}}), \eprint{2008.00501}.

\bibitem[{\citenamefont{Bolognino et~al.}(2021{\natexlab{b}})}]{Bolognino:2021mrc}
\bibinfo{author}{\bibfnamefont{A.~D.} \bibnamefont{Bolognino}} \bibnamefont{et~al.}, \bibinfo{journal}{Phys. Rev. D} \textbf{\bibinfo{volume}{103}}, \bibinfo{pages}{094004} (\bibinfo{year}{2021}{\natexlab{b}}), \eprint{2103.07396}.

\bibitem[{\citenamefont{van Hameren et~al.}(2022)\citenamefont{van Hameren, Motyka,  Ziarko}}]{vanHameren:2022mtk}
\bibinfo{author}{\bibfnamefont{A.}~\bibnamefont{van Hameren}}, \bibinfo{author}{\bibfnamefont{L.}~\bibnamefont{Motyka}},  \bibinfo{author}{\bibfnamefont{G.}~\bibnamefont{Ziarko}}, \bibinfo{journal}{JHEP} \textbf{\bibinfo{volume}{11}}, \bibinfo{pages}{103} (\bibinfo{year}{2022}), \eprint{2205.09585}.

\bibitem[{\citenamefont{Bonvini\mbox{, S. Marzani}}(2018)}]{Bonvini:2018ixe}
\bibinfo{author}{\bibfnamefont{M.}~\bibnamefont{Bonvini\mbox{, S. Marzani}}}, \bibinfo{journal}{Phys. Rev. Lett.} \textbf{\bibinfo{volume}{120}}, \bibinfo{pages}{202003} (\bibinfo{year}{2018}), \eprint{1802.07758}.

\bibitem[{\citenamefont{Silvetti\mbox{, M. Bonvini}}(2023)}]{Silvetti:2022hyc}
\bibinfo{author}{\bibfnamefont{F.}~\bibnamefont{Silvetti\mbox{, M. Bonvini}}}, \bibinfo{journal}{Eur. Phys. J. C} \textbf{\bibinfo{volume}{83}}, \bibinfo{pages}{267} (\bibinfo{year}{2023}), \eprint{2211.10142}.

\bibitem[{\citenamefont{Duclou\'e et~al.}(2013)\citenamefont{Duclou\'e, Szymanowski,  Wallon}}]{Ducloue:2013hia}
\bibinfo{author}{\bibfnamefont{B.}~\bibnamefont{Duclou\'e}}, \bibinfo{author}{\bibfnamefont{L.}~\bibnamefont{Szymanowski}},  \bibinfo{author}{\bibfnamefont{S.}~\bibnamefont{Wallon}}, \bibinfo{journal}{JHEP} \textbf{\bibinfo{volume}{05}}, \bibinfo{pages}{096} (\bibinfo{year}{2013}), \eprint{1302.7012}.

\bibitem[{\citenamefont{Duclou\'e et~al.}(2014)\citenamefont{Duclou\'e, Szymanowski,  Wallon}}]{Ducloue:2013bva}
\bibinfo{author}{\bibfnamefont{B.}~\bibnamefont{Duclou\'e}}, \bibinfo{author}{\bibfnamefont{L.}~\bibnamefont{Szymanowski}},  \bibinfo{author}{\bibfnamefont{S.}~\bibnamefont{Wallon}}, \bibinfo{journal}{Phys. Rev. Lett.} \textbf{\bibinfo{volume}{112}}, \bibinfo{pages}{082003} (\bibinfo{year}{2014}), \eprint{1309.3229}.

\bibitem[{\citenamefont{Celiberto et~al.}(2015{\natexlab{a}})}]{Celiberto:2015yba}
\bibinfo{author}{\bibfnamefont{F.~G.} \bibnamefont{Celiberto}} \bibnamefont{et~al.}, \bibinfo{journal}{Eur. Phys. J. C} \textbf{\bibinfo{volume}{75}}, \bibinfo{pages}{292} (\bibinfo{year}{2015}{\natexlab{a}}), \eprint{1504.08233}.

\bibitem[{\citenamefont{Celiberto et~al.}(2015{\natexlab{b}})}]{Celiberto:2015mpa}
\bibinfo{author}{\bibfnamefont{F.~G.} \bibnamefont{Celiberto}} \bibnamefont{et~al.}, \bibinfo{journal}{Acta Phys. Polon. Supp.} \textbf{\bibinfo{volume}{8}}, \bibinfo{pages}{935} (\bibinfo{year}{2015}{\natexlab{b}}), \eprint{1510.01626}.

\bibitem[{\citenamefont{Celiberto et~al.}(2016{\natexlab{a}})}]{Celiberto:2016ygs}
\bibinfo{author}{\bibfnamefont{F.~G.} \bibnamefont{Celiberto}} \bibnamefont{et~al.}, \bibinfo{journal}{Eur. Phys. J. C} \textbf{\bibinfo{volume}{76}}, \bibinfo{pages}{224} (\bibinfo{year}{2016}{\natexlab{a}}), \eprint{1601.07847}.

\bibitem[{\citenamefont{Celiberto}(2017)}]{Celiberto:2017ius}
\bibinfo{author}{\bibfnamefont{F.~G.} \bibnamefont{Celiberto}}, Ph.D. thesis (\bibinfo{year}{2017}), \eprint{1707.04315}.

\bibitem[{\citenamefont{Caporale et~al.}(2018)}]{Caporale:2018qnm}
\bibinfo{author}{\bibfnamefont{F.}~\bibnamefont{Caporale}} \bibnamefont{et~al.}, \bibinfo{journal}{Nucl. Phys. B} \textbf{\bibinfo{volume}{935}}, \bibinfo{pages}{412} (\bibinfo{year}{2018}), \eprint{1806.06309}.

\bibitem[{\citenamefont{Celiberto{, A. Papa}}(2022)}]{Celiberto:2022gji}
\bibinfo{author}{\bibfnamefont{F.~G.} \bibnamefont{Celiberto{, A. Papa}}}, \bibinfo{journal}{Phys. Rev. D} \textbf{\bibinfo{volume}{106}}, \bibinfo{pages}{114004} (\bibinfo{year}{2022}), \eprint{2207.05015}.

\bibitem[{\citenamefont{Egorov{, V. T. Kim}}(2023)}]{Egorov:2023duz}
\bibinfo{author}{\bibfnamefont{A.~I.} \bibnamefont{Egorov{, V. T. Kim}}}, \bibinfo{journal}{Phys. Rev. D} \textbf{\bibinfo{volume}{108}}, \bibinfo{pages}{014010} (\bibinfo{year}{2023}), \eprint{2305.19854}.

\bibitem[{\citenamefont{Baldenegro et~al.}(2024)}]{Baldenegro:2024ndr}
\bibinfo{author}{\bibfnamefont{C.}~\bibnamefont{Baldenegro}} \bibnamefont{et~al.} (\bibinfo{year}{2024}), \eprint{2406.10681}.

\bibitem[{\citenamefont{Celiberto et~al.}(2018{\natexlab{a}})}]{Celiberto:2018muu}
\bibinfo{author}{\bibfnamefont{F.~G.} \bibnamefont{Celiberto}} \bibnamefont{et~al.}, \bibinfo{journal}{Phys. Lett.} \textbf{\bibinfo{volume}{B786}}, \bibinfo{pages}{201} (\bibinfo{year}{2018}{\natexlab{a}}), \eprint{1808.09511}.

\bibitem[{\citenamefont{Golec-Biernat et~al.}(2018)}]{Golec-Biernat:2018kem}
\bibinfo{author}{\bibfnamefont{K.}~\bibnamefont{Golec-Biernat}} \bibnamefont{et~al.}, \bibinfo{journal}{JHEP} \textbf{\bibinfo{volume}{12}}, \bibinfo{pages}{091} (\bibinfo{year}{2018}), \eprint{1811.04361}.

\bibitem[{\citenamefont{Celiberto et~al.}(2016{\natexlab{b}})}]{Celiberto:2016hae}
\bibinfo{author}{\bibfnamefont{F.~G.} \bibnamefont{Celiberto}} \bibnamefont{et~al.}, \bibinfo{journal}{Phys. Rev. D} \textbf{\bibinfo{volume}{94}}, \bibinfo{pages}{034013} (\bibinfo{year}{2016}{\natexlab{b}}), \eprint{1604.08013}.

\bibitem[{\citenamefont{Celiberto et~al.}(2017)}]{Celiberto:2017ptm}
\bibinfo{author}{\bibfnamefont{F.~G.} \bibnamefont{Celiberto}} \bibnamefont{et~al.}, \bibinfo{journal}{Eur. Phys. J. C} \textbf{\bibinfo{volume}{77}}, \bibinfo{pages}{382} (\bibinfo{year}{2017}), \eprint{1701.05077}.

\bibitem[{\citenamefont{Bolognino et~al.}(2018{\natexlab{b}})}]{Bolognino:2018oth}
\bibinfo{author}{\bibfnamefont{A.~D.} \bibnamefont{Bolognino}} \bibnamefont{et~al.}, \bibinfo{journal}{Eur. Phys. J. C} \textbf{\bibinfo{volume}{78}}, \bibinfo{pages}{772} (\bibinfo{year}{2018}{\natexlab{b}}), \eprint{1808.05483}.

\bibitem[{\citenamefont{Bolognino et~al.}(2019{\natexlab{a}})}]{Bolognino:2019yqj}
\bibinfo{author}{\bibfnamefont{A.~D.} \bibnamefont{Bolognino}} \bibnamefont{et~al.}, \bibinfo{journal}{Acta Phys. Polon. Supp.} \textbf{\bibinfo{volume}{12}}, \bibinfo{pages}{773} (\bibinfo{year}{2019}{\natexlab{a}}), \eprint{1902.04511}.

\bibitem[{\citenamefont{Bolognino et~al.}(2019{\natexlab{b}})}]{Bolognino:2019cac}
\bibinfo{author}{\bibfnamefont{A.~D.} \bibnamefont{Bolognino}} \bibnamefont{et~al.}, \bibinfo{journal}{PoS} \textbf{\bibinfo{volume}{DIS2019}}, \bibinfo{pages}{049} (\bibinfo{year}{2019}{\natexlab{b}}), \eprint{1906.11800}.

\bibitem[{\citenamefont{Celiberto et~al.}(2020)\citenamefont{Celiberto, Ivanov,  Papa}}]{Celiberto:2020rxb}
\bibinfo{author}{\bibfnamefont{F.~G.} \bibnamefont{Celiberto}}, \bibinfo{author}{\bibfnamefont{D.~{\relax Yu}.} \bibnamefont{Ivanov}},  \bibinfo{author}{\bibfnamefont{A.}~\bibnamefont{Papa}}, \bibinfo{journal}{Phys. Rev. D} \textbf{\bibinfo{volume}{102}}, \bibinfo{pages}{094019} (\bibinfo{year}{2020}), \eprint{2008.10513}.

\bibitem[{\citenamefont{Celiberto}(2023{\natexlab{a}})}]{Celiberto:2022kxx}
\bibinfo{author}{\bibfnamefont{F.~G.} \bibnamefont{Celiberto}}, \bibinfo{journal}{Eur. Phys. J. C} \textbf{\bibinfo{volume}{83}}, \bibinfo{pages}{332} (\bibinfo{year}{2023}{\natexlab{a}}), \eprint{2208.14577}.

\bibitem[{\citenamefont{Celiberto et~al.}(2018{\natexlab{b}})}]{Celiberto:2017nyx}
\bibinfo{author}{\bibfnamefont{F.~G.} \bibnamefont{Celiberto}} \bibnamefont{et~al.}, \bibinfo{journal}{Phys. Lett. B} \textbf{\bibinfo{volume}{777}}, \bibinfo{pages}{141} (\bibinfo{year}{2018}{\natexlab{b}}), \eprint{1709.10032}.

\bibitem[{\citenamefont{Bolognino et~al.}(2019{\natexlab{c}})}]{Bolognino:2019yls}
\bibinfo{author}{\bibfnamefont{A.~D.} \bibnamefont{Bolognino}} \bibnamefont{et~al.}, \bibinfo{journal}{Eur. Phys. J. C} \textbf{\bibinfo{volume}{79}}, \bibinfo{pages}{939} (\bibinfo{year}{2019}{\natexlab{c}}), \eprint{1909.03068}.

\bibitem[{\citenamefont{Bolognino et~al.}(2019{\natexlab{d}})}]{Bolognino:2019ccd}
\bibinfo{author}{\bibfnamefont{A.~D.} \bibnamefont{Bolognino}} \bibnamefont{et~al.}, \bibinfo{journal}{PoS} \textbf{\bibinfo{volume}{DIS2019}}, \bibinfo{pages}{067} (\bibinfo{year}{2019}{\natexlab{d}}), \eprint{1906.05940}.

\bibitem[{\citenamefont{Adachi et~al.}(2022)}]{AlexanderAryshev:2022pkx}
\bibinfo{author}{\bibfnamefont{I.}~\bibnamefont{Adachi}} \bibnamefont{et~al.} (\bibinfo{collaboration}{ILC International Community}) (\bibinfo{year}{2022}), \eprint{2203.07622}.

\bibitem[{\citenamefont{Celiberto et~al.}(2021{\natexlab{b}})}]{Celiberto:2021dzy}
\bibinfo{author}{\bibfnamefont{F.~G.} \bibnamefont{Celiberto}} \bibnamefont{et~al.}, \bibinfo{journal}{Eur. Phys. J. C} \textbf{\bibinfo{volume}{81}}, \bibinfo{pages}{780} (\bibinfo{year}{2021}{\natexlab{b}}), \eprint{2105.06432}.

\bibitem[{\citenamefont{Celiberto et~al.}(2021{\natexlab{c}})}]{Celiberto:2021fdp}
\bibinfo{author}{\bibfnamefont{F.~G.} \bibnamefont{Celiberto}} \bibnamefont{et~al.}, \bibinfo{journal}{Phys. Rev. D} \textbf{\bibinfo{volume}{104}}, \bibinfo{pages}{114007} (\bibinfo{year}{2021}{\natexlab{c}}), \eprint{2109.11875}.

\bibitem[{\citenamefont{Celiberto et~al.}(2022{\natexlab{a}})}]{Celiberto:2022zdg}
\bibinfo{author}{\bibfnamefont{F.~G.} \bibnamefont{Celiberto}} \bibnamefont{et~al.}, \bibinfo{journal}{Phys. Rev. D} \textbf{\bibinfo{volume}{105}}, \bibinfo{pages}{114056} (\bibinfo{year}{2022}{\natexlab{a}}), \eprint{2205.13429}.

\bibitem[{\citenamefont{Celiberto}(2022{\natexlab{a}})}]{Celiberto:2022keu}
\bibinfo{author}{\bibfnamefont{F.~G.} \bibnamefont{Celiberto}}, \bibinfo{journal}{Phys. Lett. B} \textbf{\bibinfo{volume}{835}}, \bibinfo{pages}{137554} (\bibinfo{year}{2022}{\natexlab{a}}), \eprint{2206.09413}.

\bibitem[{\citenamefont{Celiberto}(2024{\natexlab{a}})}]{Celiberto:2024omj}
\bibinfo{author}{\bibfnamefont{F.~G.} \bibnamefont{Celiberto}}, \bibinfo{journal}{Eur. Phys. J. C} \textbf{\bibinfo{volume}{84}}, \bibinfo{pages}{384} (\bibinfo{year}{2024}{\natexlab{a}}), \eprint{2401.01410}.

\bibitem[{\citenamefont{Anchordoqui et~al.}(2022)}]{Anchordoqui:2021ghd}
\bibinfo{author}{\bibfnamefont{L.~A.} \bibnamefont{Anchordoqui}} \bibnamefont{et~al.}, \bibinfo{journal}{Phys. Rept.} \textbf{\bibinfo{volume}{968}}, \bibinfo{pages}{1} (\bibinfo{year}{2022}), \eprint{2109.10905}.

\bibitem[{\citenamefont{Feng et~al.}(2023)}]{Feng:2022inv}
\bibinfo{author}{\bibfnamefont{J.~L.} \bibnamefont{Feng}} \bibnamefont{et~al.}, \bibinfo{journal}{J. Phys. G} \textbf{\bibinfo{volume}{50}}, \bibinfo{pages}{030501} (\bibinfo{year}{2023}), \eprint{2203.05090}.

\bibitem[{\citenamefont{Boussarie et~al.}(2018)}]{Boussarie:2017oae}
\bibinfo{author}{\bibfnamefont{R.}~\bibnamefont{Boussarie}} \bibnamefont{et~al.}, \bibinfo{journal}{Phys. Rev. D} \textbf{\bibinfo{volume}{97}}, \bibinfo{pages}{014008} (\bibinfo{year}{2018}), \eprint{1709.01380}.

\bibitem[{\citenamefont{Chapon et~al.}(2022)}]{Chapon:2020heu}
\bibinfo{author}{\bibfnamefont{E.}~\bibnamefont{Chapon}} \bibnamefont{et~al.}, \bibinfo{journal}{Prog. Part. Nucl. Phys.} \textbf{\bibinfo{volume}{122}}, \bibinfo{pages}{103906} (\bibinfo{year}{2022}), \eprint{2012.14161}.

\bibitem[{\citenamefont{Celiberto\mbox{, M. Fucilla}}(2022)}]{Celiberto:2022dyf}
\bibinfo{author}{\bibfnamefont{F.~G.} \bibnamefont{Celiberto\mbox{, M. Fucilla}}}, \bibinfo{journal}{Eur. Phys. J. C} \textbf{\bibinfo{volume}{82}}, \bibinfo{pages}{929} (\bibinfo{year}{2022}), \eprint{2202.12227}.

\bibitem[{\citenamefont{Celiberto}(2023{\natexlab{b}})}]{Celiberto:2023fzz}
\bibinfo{author}{\bibfnamefont{F.~G.} \bibnamefont{Celiberto}}, \bibinfo{journal}{Universe} \textbf{\bibinfo{volume}{9}}, \bibinfo{pages}{324} (\bibinfo{year}{2023}{\natexlab{b}}), \eprint{2305.14295}.

\bibitem[{\citenamefont{Celiberto\mbox{, A. Papa}}(2024)}]{Celiberto:2023rzw}
\bibinfo{author}{\bibfnamefont{F.~G.} \bibnamefont{Celiberto\mbox{, A. Papa}}}, \bibinfo{journal}{Phys. Lett. B} \textbf{\bibinfo{volume}{848}}, \bibinfo{pages}{138406} (\bibinfo{year}{2024}), \eprint{2308.00809}.

\bibitem[{\citenamefont{Celiberto et~al.}(2024{\natexlab{a}})\citenamefont{Celiberto, Gatto,  Papa}}]{Celiberto:2024mab}
\bibinfo{author}{\bibfnamefont{F.~G.} \bibnamefont{Celiberto}}, \bibinfo{author}{\bibfnamefont{G.}~\bibnamefont{Gatto}},  \bibinfo{author}{\bibfnamefont{A.}~\bibnamefont{Papa}} (\bibinfo{year}{2024}{\natexlab{a}}), \eprint{2405.14773}.

\bibitem[{\citenamefont{Celiberto}(2024{\natexlab{b}})}]{Celiberto:2024mrq}
\bibinfo{author}{\bibfnamefont{F.~G.} \bibnamefont{Celiberto}}, \bibinfo{journal}{Symmetry} \textbf{\bibinfo{volume}{16}}, \bibinfo{pages}{550} (\bibinfo{year}{2024}{\natexlab{b}}), \eprint{2403.15639}.

\bibitem[{\citenamefont{Celiberto\mbox{, A. Papa}}(2023)}]{Celiberto:2023rtu}
\bibinfo{author}{\bibfnamefont{F.~G.} \bibnamefont{Celiberto\mbox{, A. Papa}}} (\bibinfo{year}{2023}), \eprint{2305.00962}.

\bibitem[{\citenamefont{Chen et~al.}(2015)}]{Chen:2014gva}
\bibinfo{author}{\bibfnamefont{X.}~\bibnamefont{Chen}} \bibnamefont{et~al.}, \bibinfo{journal}{Phys. Lett. B} \textbf{\bibinfo{volume}{740}}, \bibinfo{pages}{147} (\bibinfo{year}{2015}), \eprint{1408.5325}.

\bibitem[{\citenamefont{Boughezal et~al.}(2015)}]{Boughezal:2015dra}
\bibinfo{author}{\bibfnamefont{R.}~\bibnamefont{Boughezal}} \bibnamefont{et~al.}, \bibinfo{journal}{Phys. Rev. Lett.} \textbf{\bibinfo{volume}{115}}, \bibinfo{pages}{082003} (\bibinfo{year}{2015}), \eprint{1504.07922}.

\bibitem[{\citenamefont{Dawson et~al.}(2022)}]{Dawson:2022zbb}
\bibinfo{author}{\bibfnamefont{S.}~\bibnamefont{Dawson}} \bibnamefont{et~al.} (\bibinfo{year}{2022}), \eprint{2209.07510}.

\bibitem[{\citenamefont{Monni et~al.}(2020)\citenamefont{Monni, Rottoli,  Torrielli}}]{Monni:2019yyr}
\bibinfo{author}{\bibfnamefont{P.~F.} \bibnamefont{Monni}}, \bibinfo{author}{\bibfnamefont{L.}~\bibnamefont{Rottoli}},  \bibinfo{author}{\bibfnamefont{P.}~\bibnamefont{Torrielli}}, \bibinfo{journal}{Phys. Rev. Lett.} \textbf{\bibinfo{volume}{124}}, \bibinfo{pages}{252001} (\bibinfo{year}{2020}), \eprint{1909.04704}.

\bibitem[{\citenamefont{Hentschinski et~al.}(2021)}]{Hentschinski:2020tbi}
\bibinfo{author}{\bibfnamefont{M.}~\bibnamefont{Hentschinski}} \bibnamefont{et~al.}, \bibinfo{journal}{Eur. Phys. J. C} \textbf{\bibinfo{volume}{81}}, \bibinfo{pages}{112} (\bibinfo{year}{2021}), \eprint{2011.03193}.

\bibitem[{\citenamefont{Celiberto et~al.}(2022{\natexlab{b}})}]{Celiberto:2022fgx}
\bibinfo{author}{\bibfnamefont{F.~G.} \bibnamefont{Celiberto}} \bibnamefont{et~al.}, \bibinfo{journal}{JHEP} \textbf{\bibinfo{volume}{08}}, \bibinfo{pages}{092} (\bibinfo{year}{2022}{\natexlab{b}}), \eprint{2205.02681}.

\bibitem[{\citenamefont{Nefedov}(2019)}]{Nefedov:2019mrg}
\bibinfo{author}{\bibfnamefont{M.~A.} \bibnamefont{Nefedov}}, \bibinfo{journal}{Nucl. Phys. B} \textbf{\bibinfo{volume}{946}}, \bibinfo{pages}{114715} (\bibinfo{year}{2019}), \eprint{1902.11030}.

\bibitem[{\citenamefont{Celiberto}(2021{\natexlab{b}})}]{Celiberto:2020wpk}
\bibinfo{author}{\bibfnamefont{F.~G.} \bibnamefont{Celiberto}}, \bibinfo{journal}{Eur. Phys. J. C} \textbf{\bibinfo{volume}{81}}, \bibinfo{pages}{691} (\bibinfo{year}{2021}{\natexlab{b}}), \eprint{2008.07378}.

\bibitem[{\citenamefont{Celiberto}(2022{\natexlab{b}})}]{Celiberto:2022rfj}
\bibinfo{author}{\bibfnamefont{F.~G.} \bibnamefont{Celiberto}}, \bibinfo{journal}{Phys. Rev. D} \textbf{\bibinfo{volume}{105}}, \bibinfo{pages}{114008} (\bibinfo{year}{2022}{\natexlab{b}}), \eprint{2204.06497}.

\bibitem[{\citenamefont{Celiberto}(2024{\natexlab{c}})}]{Celiberto:2024swu}
\bibinfo{author}{\bibfnamefont{F.~G.} \bibnamefont{Celiberto}}, \bibinfo{journal}{Particles} \textbf{\bibinfo{volume}{7}}, \bibinfo{pages}{502} (\bibinfo{year}{2024}{\natexlab{c}}), \eprint{2405.09526}.

\bibitem[{\citenamefont{Celiberto et~al.}(2023)}]{Celiberto:2023uuk_article}
\bibinfo{author}{\bibfnamefont{F.~G.} \bibnamefont{Celiberto}} \bibnamefont{et~al.}, \bibinfo{journal}{\emph{Proceedings of Moriond QCD}}  (\bibinfo{year}{2023}), \eprint{2305.05052}.

\bibitem[{\citenamefont{Celiberto et~al.}(2024{\natexlab{b}})}]{Celiberto:2023eba_article}
\bibinfo{author}{\bibfnamefont{F.~G.} \bibnamefont{Celiberto}} \bibnamefont{et~al.}, \bibinfo{journal}{PoS} \textbf{\bibinfo{volume}{RADCOR2023}}, \bibinfo{pages}{069} (\bibinfo{year}{2024}{\natexlab{b}}), \eprint{2309.11573}.

\bibitem[{\citenamefont{Celiberto et~al.}(2024{\natexlab{c}})}]{Celiberto:2023nym}
\bibinfo{author}{\bibfnamefont{F.~G.} \bibnamefont{Celiberto}} \bibnamefont{et~al.}, \bibinfo{journal}{PoS} \textbf{\bibinfo{volume}{EPS-HEP2023}}, \bibinfo{pages}{390} (\bibinfo{year}{2024}{\natexlab{c}}), \eprint{2310.16967}.

\bibitem[{\citenamefont{Hamilton et~al.}(2013)}]{Hamilton:2012rf}
\bibinfo{author}{\bibfnamefont{K.}~\bibnamefont{Hamilton}} \bibnamefont{et~al.}, \bibinfo{journal}{JHEP} \textbf{\bibinfo{volume}{05}}, \bibinfo{pages}{082} (\bibinfo{year}{2013}), \eprint{1212.4504}.

\bibitem[{\citenamefont{Bagnaschi et~al.}(2023)}]{Bagnaschi:2023rbx}
\bibinfo{author}{\bibfnamefont{E.}~\bibnamefont{Bagnaschi}} \bibnamefont{et~al.}, \bibinfo{journal}{Eur. Phys. J. C} \textbf{\bibinfo{volume}{83}}, \bibinfo{pages}{1054} (\bibinfo{year}{2023}), \eprint{2309.10525}.

\bibitem[{\citenamefont{Banfi et~al.}(2024)}]{Banfi:2023mhz}
\bibinfo{author}{\bibfnamefont{A.}~\bibnamefont{Banfi}} \bibnamefont{et~al.}, \bibinfo{journal}{JHEP} \textbf{\bibinfo{volume}{02}}, \bibinfo{pages}{023} (\bibinfo{year}{2024}), \eprint{2309.02127}.

\bibitem[{\citenamefont{Alioli et~al.}(2023)}]{Alioli:2022dkj}
\bibinfo{author}{\bibfnamefont{S.}~\bibnamefont{Alioli}} \bibnamefont{et~al.}, \bibinfo{journal}{JHEP} \textbf{\bibinfo{volume}{06}}, \bibinfo{pages}{205} (\bibinfo{year}{2023}), \eprint{2212.10489}.

\bibitem[{\citenamefont{van Beekveld et~al.}(2022)}]{vanBeekveld:2022zhl}
\bibinfo{author}{\bibfnamefont{M.}~\bibnamefont{van Beekveld}} \bibnamefont{et~al.}, \bibinfo{journal}{JHEP} \textbf{\bibinfo{volume}{11}}, \bibinfo{pages}{019} (\bibinfo{year}{2022}), \eprint{2205.02237}.

\bibitem[{\citenamefont{Ferrario~Ravasio et~al.}(2023)}]{FerrarioRavasio:2023kyg}
\bibinfo{author}{\bibfnamefont{S.}~\bibnamefont{Ferrario~Ravasio}} \bibnamefont{et~al.}, \bibinfo{journal}{Phys. Rev. Lett.} \textbf{\bibinfo{volume}{131}}, \bibinfo{pages}{161906} (\bibinfo{year}{2023}), \eprint{2307.11142}.

\bibitem[{\citenamefont{Jones et~al.}(2018)\citenamefont{Jones, Kerner,  Luisoni}}]{Jones:2018hbb}
\bibinfo{author}{\bibfnamefont{S.~P.} \bibnamefont{Jones}}, \bibinfo{author}{\bibfnamefont{M.}~\bibnamefont{Kerner}},  \bibinfo{author}{\bibfnamefont{G.}~\bibnamefont{Luisoni}}, \bibinfo{journal}{Phys. Rev. Lett.} \textbf{\bibinfo{volume}{120}}, \bibinfo{pages}{162001} (\bibinfo{year}{2018}), \eprint{1802.00349}.

\bibitem[{\citenamefont{Bonciani et~al.}(2023)}]{Bonciani:2022jmb}
\bibinfo{author}{\bibfnamefont{R.}~\bibnamefont{Bonciani}} \bibnamefont{et~al.}, \bibinfo{journal}{Phys. Lett. B} \textbf{\bibinfo{volume}{843}}, \bibinfo{pages}{137995} (\bibinfo{year}{2023}), \eprint{2206.10490}.

\end{thebibliography}
\endgroup
%-----------------------------------------

\end{document}